\documentclass[letter]{ptptex}

\usepackage{graphicx}
\usepackage{subfigure}

\notypesetlogo                       
\preprintnumber[3cm]{
KUNS-1325}

\title{
Negative Differential Resistivity from Holography%
}

\author{
Shin \textsc{Nakamura}%
}

\inst{
Department of Physics, Kyoto University, Kyoto 606-8502, Japan
}

\recdate{September 22, 2010}

\abst{
Negative differential resistivity (NDR) in a (3+1)-dimensional quantum system of strongly correlated charge carriers is theoretically reproduced by using the AdS/CFT correspondence. Our system is microscopically defined, and the analysis does not rely on any phenomenological models of NDR mechanism. The interaction between the heat reservoir and the nonequilibrium charge carriers is also taken into account. The system realizes insulation due to the formation of the neutral bound states of the charge carriers  at low temperatures. However, the insulation is broken and the field-induced conductor phase appears by applying the external electric field greater than the critical value. We find that the NDR appears right above the critical electric field in this phase, and the pair creation of the charge carriers is crucial to realize the NDR. The present result suggests a possibility to observe a similar NDR in excitonic insulators or in the strongly interacting quark-gluon plasma, where the insulation originates in the formation of the neutral bound states.
}

\begin{document}

\maketitle

\begin{figure}[h]
\centering
\subfigure[S-shaped NDR]{
\includegraphics[keepaspectratio=true,height=40mm]{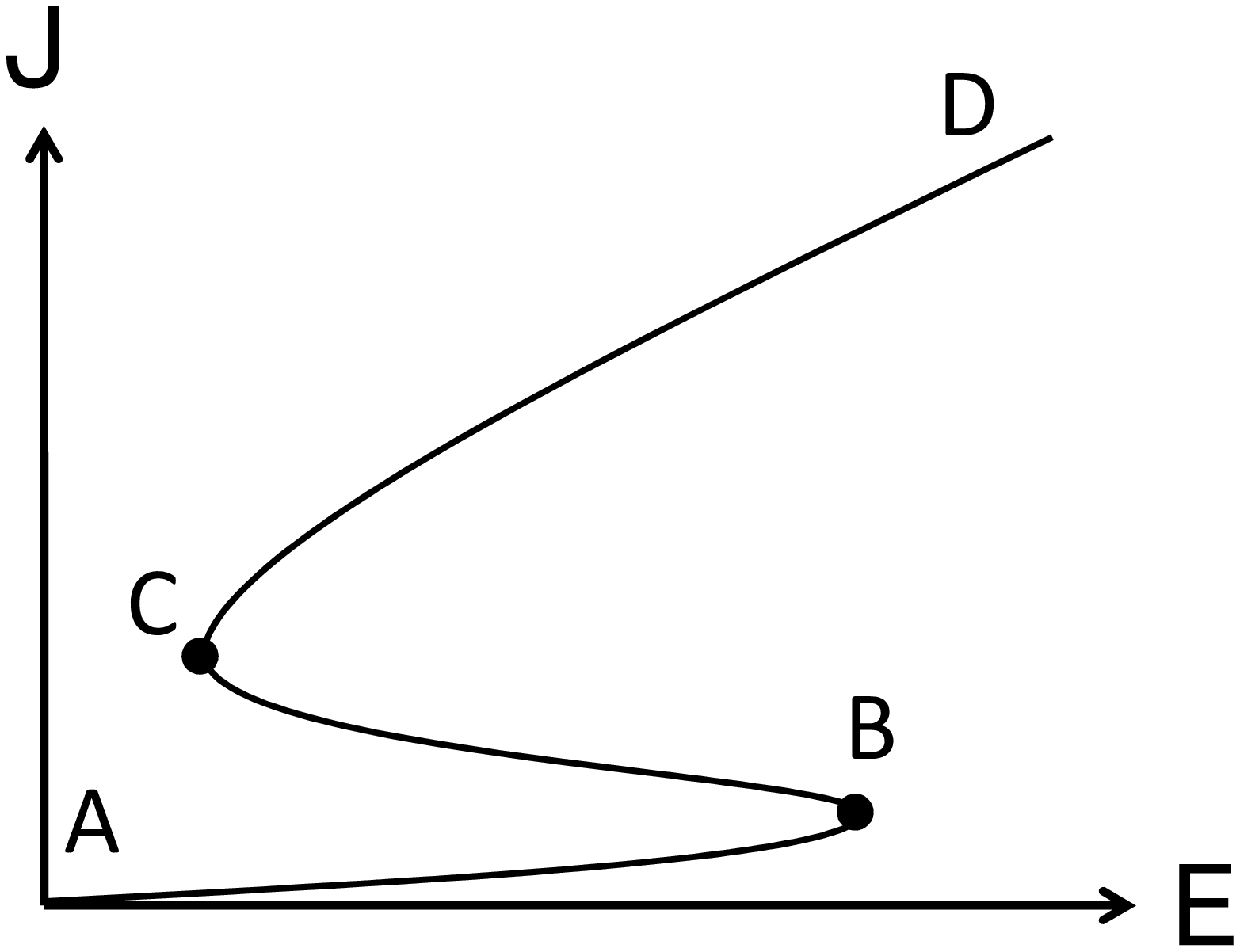}
\label{fig:S-shape}}
\subfigure[N-shaped NDR]{
\includegraphics[keepaspectratio=true,height=40mm]{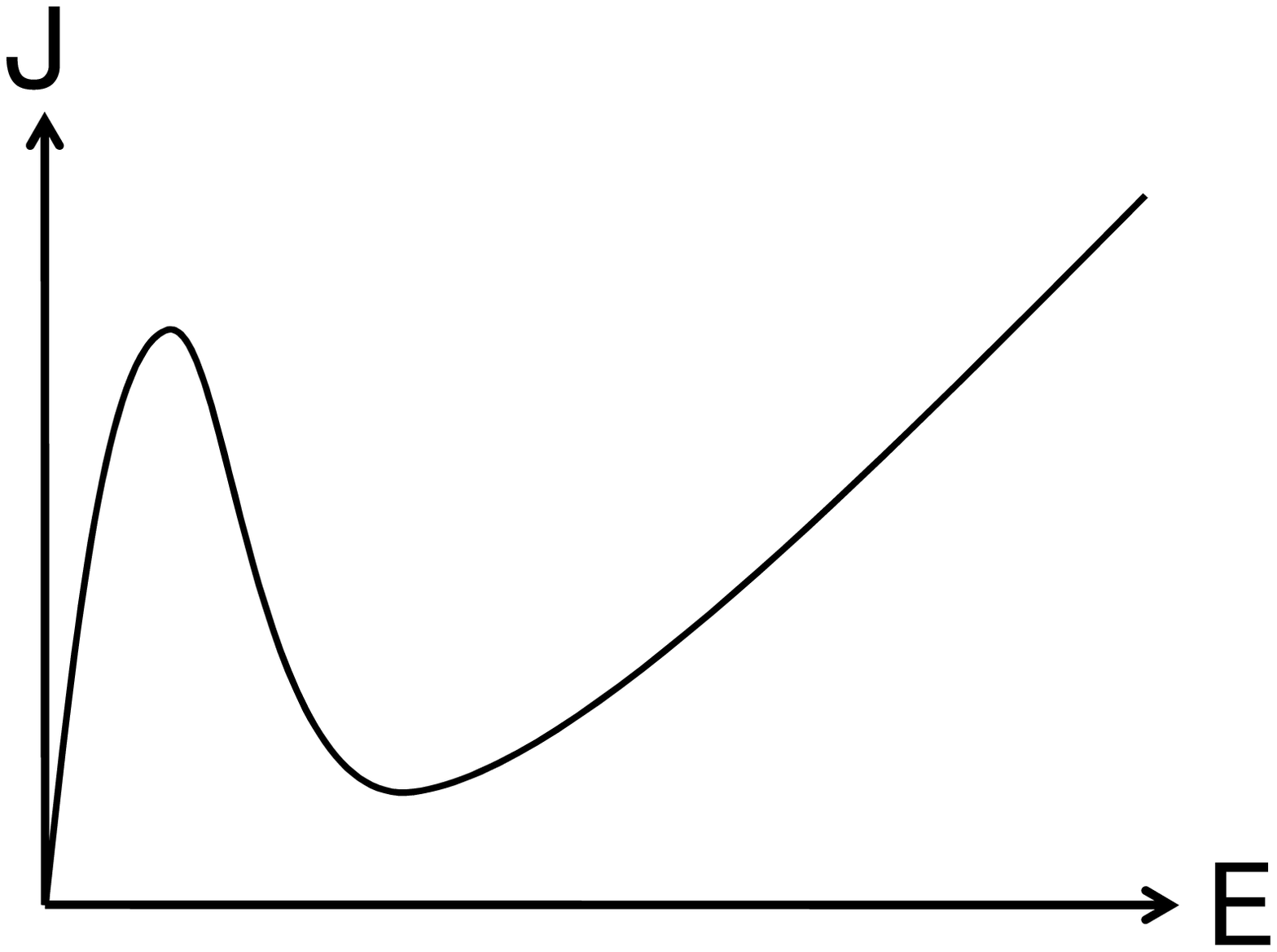}
\label{fig:N-shape}}
\caption{Typical $J$-$E$ characteristics with NDR (schematic).}
\label{fig:NDR}
\end{figure}

{\bf Introduction.}---Nonlinear electric conduction in strongly correlated electron systems (SCESs), such as Mott and charge order insulators, is characterized by their threshold property (see, for example, Refs.~\citen{SLZ})
and negative differential resistivity (NDR)\footnote{This may also be referred to as negative differential conductivity (NDC) in some literature.}. NDR is a nonlinear phenomenon in charge transport where the electric field ($E$) decreases with increasing current density ($J$), and vice versa (Fig.~\ref{fig:NDR}). This has been observed in various materials and devices\cite{book}, and one category of such materials is SCES (see, for example, Refs.~\citen{exp1,exp1-2,exp2}). 
Since electronic devices that exhibit NDR are useful in electric circuits, the understanding of NDR is also important from the viewpoint of industrial applications. 
However, as far as the author knows, most of the theoretical studies on NDR are based on the phenomenological models of NDR mechanism~\cite{book}, except for some (quasi) one-dimensional systems~\cite{Ajisaka2009,Beneti2009xxz,Beneti2009hubbard,Koutouza2000wire}.\footnote{The NDR in Esaki diode~\cite{Esaki}, for example, is explained by the tunnel effect of the charge carriers at the p-n junctions of semiconductors. However, we consider only uniform materials 
in this article.}

The theoretical difficulties in the study of NDR in SCES come from the following facts:
\begin{itemize}
  \item NDR is a nonlinear phenomenon where we need to go beyond the linear response theory.
  \item This also means that the system is far from equilibrium owing to the dissipation caused by the finite current.
  \item Nonperturbative analysis is necessary if the NDR is associated with the threshold property such as the metal-insulator transition, a transition of the vacuum.
  \item Nonperturbative analysis may also be needed in some SCES owing to the sufficiently strong effective interaction.\footnote{For example, the effective fine structure constant in graphene is $O(1)$ since the Fermi velocity of the electrons is about 300 times slower than the speed  of light.\cite{graphene} (See also Ref.~\citen{Araki:2010gj}.)}
\end{itemize}
In this article, we apply the AdS/CFT correspondence 
(or the gauge/gravity duality, or holography)~\cite{Maldacena:1997re,Gubser:1998bc,Witten:1998qj} to overcome these difficulties, and we reproduce NDR without relying on any phenomenological modeling of the NDR mechanism. 
The AdS/CFT correspondence is a correspondence between strongly interacting quantum gauge theories and higher-dimensional classical gravities. This enables us to analyze the system nonperturbatively, and we can compute the nonlinear conductivity~\cite{Karch:2007pd} of a global charge.
These are remarkable advantages of the AdS/CFT correspondence. 
We consider an idealized gauge theory whose gravity dual is well-established and that shares several features similar to those of the excitonic insulators~\cite{excitonic,excitonic-review} or sQGP~\cite{Gyulassy:2004zy}. 
We find that the system shows NDR due to the pair-creation process of the charge carriers. Our result suggests a possibility to observe NDR in some excitonic insulators or in some quark-hadron systems, as we shall discuss later.

The typical $J$-$E$ characteristics with NDR are divided into two categories\cite{book}: the S-shaped NDR (Fig.~\ref{fig:S-shape}) and the N-shaped NDR (Fig.~\ref{fig:N-shape}).\footnote{They correspond to the SNDC and NNDC in Ref.~\citen{book}, respectively.} 
NDR is realized between B and C in Fig.~\ref{fig:S-shape} and is realized between the local maximum and the local minimum in Fig.~\ref{fig:N-shape}. The famous NDR behavior of the Esaki diode~\cite{Esaki} is the N-shaped NDR, while the typical $J$-$E$ characteristics of SCES are the S-shaped NDR (see, for example, Refs.~\citen{exp1,exp1-2,exp2}).
In the S-shaped NDR, $J(E)$ is a multivalued function of $E$. Experimentally, the multivalued behavior is obtained by measuring $E$ as a function of the controlled current density $J$; note that the $J$-$E$ curve is ``N-shaped'' if the axes are swapped.
The function $E$ of $J$ is still a single-valued function, and the NDR is well-defined if $J$ is controlled. If we control $E$ instead, the NDR branch is unstable and the hysteresis is observed (see, for example, Ref.~\citen{exp1-2}). What we shall find in our system is the S-shaped NDR. In this article, we regard $J$ as a control parameter and $E$ is determined as a result of dynamics.

{\bf Microscopic theory.}---We employ (3+1)-dimensional $SU(N_{c})$ ${\cal N}$=4 super-symmetric Yang-Mills (SYM) theory with $N_{f}$ flavors of fundamental ${\cal N}$=2 hypermultiplets, as our microscopic theory. This is a supersymmetric cousin of Quantum Chromodynamics (QCD), but the inter-quark potential is of Coulomb type at zero temperature with infinite current quark mass due to the conformal nature of ${\cal N}=4$ SYM. The supersymmetry is broken at finite temperatures.

The number of colors $N_{c}$ is taken to be infinity with the 't Hooft coupling $g_{\mbox{\scriptsize YM}}N_{c}^{2}$ kept fixed, where $g_{\mbox{\scriptsize YM}}$ is the Yang-Mills coupling constant. 
We define $\lambda\equiv 2g_{\mbox{\scriptsize YM}}N_{c}^{2}$ in this article, and take the strong-coupling limit $\lambda \gg 1$. In terms of SCES, $\lambda$ determines the interaction between the charge carriers. In our setup, the quarks and the antiquarks are strongly correlated owing to the large $\lambda$. 
The quarks carry the global $U(1)$ baryon ($U(1)_{\mbox{\scriptsize B}}$) charge  (or the quark charge), and we analyze the conductivity associated with this charge.
We consider the deconfinement phase of gluons (which means that the degree of freedom of the gluonic sector is $O(N_{c}^{2})$), but the quark and antiquark may still form the bound states depending on the parameters of the theory. If they form the bound states, the system is an insulator since the bound states are neutral. The system becomes a conductor if the bound states are unstable and the charge carriers are liberated.  The above setup is employed to make the AdS/CFT correspondence simpler. 

Let us consider how to realize a nonequilibrium steady state (NESS) with a constant current in the conductor phase. Since our quarks/antiquarks interact strongly with the gluons,
the kinetic energy of the quarks/antiquarks will be dissipated. Because of the dissipation, the system will be heated up if we maintain a constant current. However, we can realize a steady state with a constant current by taking the probe limit $N_{c}\gg N_{f}$. The degree of freedom of the gluonic sector is $O(N_{c}^{2})$, whereas that of the flavor sector (the quark/antiquark sector) is $O(N_{c}N_{f})$; the gluonic sector has infinitely large degrees of freedom in comparison with the flavor sector at this limit. As a result, the dissipated energy from the flavor sector is absorbed into an infinitely large reservoir of gluons and the system is well-approximated as a NESS for the time period shorter than $O(N_{c})$~\cite{Karch:2008uy}. The gluonic sector acts as a ``heat bath'' for the flavor sector in this sense. Note that the interaction between the charge carriers and the ``heat bath'' is taken into account in our setup. 

{\bf Gravity dual.}---The gravity dual of the foregoing microscopic theory is the so-called D3-D7 system~\cite{Karch:2002sh}, where the $N_{f}$ D7-branes are embedded in the background geometry given by a direct product of a 5-dimensional AdS-Schwarzschild black hole (AdS-BH) and $S^{5}$.
The flavor sector is governed by the dynamics of the D7-branes, whereas the gluonic sector is described by the AdS-BH. We take the string tension to be 1 for simplicity, namely, $2\pi l_{s}^{2}=1$, where $l_{s}$ is the string length. 

The metric of the AdS-BH part is given by
\begin{eqnarray}
ds^{2}=
-\frac{1}{z^{2}}\frac{(1-z^{4}/z^{4}_{H})^{2}}{1+z^{4}/z^{4}_{H}}dt^{2}
+\frac{1+z^{4}/z^{4}_{H}}{z^{2}}d\vec{x}^{2}+\frac{dz^{2}}{z^{2}},\:\:
\label{AdS-BH}
\end{eqnarray}
where $z$ is the radial coordinate of the black hole.
The horizon is located at $z=z_{H}$ and the boundary is at $z=0$. The Hawking temperature that corresponds to the temperature of the gluonic sector (heat bath) is given by $T=\sqrt{2}/(\pi z_{H})$. $\vec{x}$ denotes the 3-dimensional spatial directions. 
The $S^{5}$ metric is
$d\Omega_{5}^{2}=d\theta^{2}+\sin^{2}\theta d\psi^{2}+\cos^{2}\theta d\Omega_{3}^{2}$,
where $0\le \theta\le \pi/2$, and $d\Omega_{d}$ is the volume element of the unit $d$-dimensional sphere. The radius of the $S^{5}$ has been taken to be 1, which is equivalent to the choice of $\lambda=(2\pi)^{2}$.

The D7-branes are wrapped on an $S^{3}$ part of the $S^{5}$. We choose our space-time coordinates in such a way that the $S^{3}$ is located at $\psi=0$.
Let us choose the worldvolume coordinates of the D7-brane to be the same as the space-time coordinates. 
We also assume that the external $U(1)_{\mbox{\scriptsize B}}$ electric field $E$ is applied along the $x$ direction. 
We have a $U(1)$ gauge field $A_{\mu}$ on the D7-branes, which couples to the $U(1)_{\mbox{\scriptsize B}}$ current. The relationship between the external field $E$ and the resulting current $J$ along the $x$ direction is given by the GKP-Witten prescription \cite{Gubser:1998bc,Witten:1998qj} as (see also Ref.~\citen{Karch:2007pd})
$A_{x}(z,t)=-Et+{\rm const.}+\frac{1}{2}\frac{J}{N}z^{2}+O(z^{4})$,
where we have employed the gauge $\partial_{x}A_{t}=0$. $N$ is given by $N=N_{f}T_{D7}(2\pi^{2})$, where $T_{D7}$
is the D7-brane tension. In our choice of $\lambda=(2\pi)^{2}$ and $2\pi l_{s}^{2}=1$, $N=N_{c}N_{f}/(2\pi)^{2}$. We consider the vanishing quark-charge density in most cases and we set the other components of the vector potential to be zero unless specified.

The D7-brane action with the present setup is explicitly written as
\begin{eqnarray}
S_{D7}=-N \int dt d^{3}x dz \cos^{3}\theta 
\Big[|g_{tt}|g_{xx}g_{zz}
-
\left(g_{zz}(\dot{A}_{x})^{2}-|g_{tt}|(A^{\prime}_{x})^{2}
\right)\Big]^{1/2},
\label{D7action}
\end{eqnarray}
where the prime (the dot) denotes the differentiation with respect to $z$ ($t$).  
We have already integrated the $S^{3}$ part under the assumption of the symmetry along it. $g_{tt}, g_{xx}$ and $g_{zz}$ are the induced world-volume metric, and they are equal to the background metric (\ref{AdS-BH}) except for $g_{zz}=1/z^{2}+\theta^{\prime}(z)^{2}$. The Wess-Zumino action of the D7-brane does not contribute in our setup.

The probe approximation is understood in the gravity dual as follows. The D7-brane action (\ref{D7action}) is affected by the AdS-BH (\ref{AdS-BH}), whereas the AdS-BH  (\ref{AdS-BH}) is not corrected by the D7-branes; the background metric (\ref{AdS-BH}) is fixed and our dynamical variables are only $A_{x}(t,z)$ and $\theta(z)$. The realization of NESS in the gravity dual is also understood in this context. If $J\neq 0$, there is a flow of energy density (which is equal to $J\cdot E$)~\cite{Karch:2008uy} along the D7-branes into the black hole horizon. Strictly speaking, the temperature of the black hole should increase owing to the energy conservation. However, the back reaction from the D7-branes to the black hole is negligible by virtue of the probe approximation, hence, the temperature is kept fixed. 

{\bf Nonlinear conductivity.}---It was found~\cite{Karch:2007pd} that the on-shell D7-brane action becomes complex unless we choose a specific combination of $J$ and $E$; the relationship between $J$ and $E$ is determined by the reality condition of the on-shell action, hence, $J$ is obtained as a nonlinear function of $E$. The on-shell action is given by~\cite{Karch:2007pd}
$\bar{S}_{D7}=-N\int dz dt d^{3}x \sqrt{\bar{g}_{zz}|g_{tt}|^{-1}}\sqrt{F_{1}F_{2}}$ with
$F_{1}=|g_{tt}|g_{xx}-E^{2}$ and
$F_{2}=|g_{tt}|g_{xx}^{2}\cos^{6}\bar{\theta}-g_{xx}J^{2}/N^{2}$,
where $\bar{g}_{zz}$ is the induced metric given by $\bar{\theta}$, which is the on-shell configuration of $\theta(z)$. Since both $F_{1}$ and $F_{2}$ cross zero somewhere between the boundary and the horizon, the only way to make $\bar{S}_{D7}$ real is to choose $J$ and $E$ so that $F_{1}$ and $F_{2}$ cross zero at the same point $z=z_{*}$. The hypersurface given by $z=z_{*}$ is often called the ``singular shell''. The location of the singular shell is given by solving $F_{1}=0$ and is found to be $z_{*}=(\sqrt{e^{2}+1}-e)^{1/2}z_{H}$, where $e=E/(\frac{\pi}{2}\sqrt{\lambda}T^{2})$. Then, the reality condition $F_{2}(z_{*})=0$ gives us the relationship between $J$ and $E$ in the form of $J=\sigma_{0}E$~\cite{Karch:2007pd}, where
\begin{equation}
\sigma_{0}=N \: T (e^{2}+1)^{1/4}\cos^{3}\bar{\theta}(z_{*}).
\label{sigma-0}
\end{equation}
We need to solve the equation of motion (EOM) for $\theta$ to obtain the explicit representation. $\bar{\theta}(z)$ can be expanded as
$\bar{\theta}(z)=m_{q}z+O(z^{3})$,
where $m_{q}$ is the current quark mass~\cite{Karch:2002sh}, which is a parameter of the microscopic theory. In terms of SCES, $m_{q}$ is related to the gap.

{\bf Qualitative behavior of D7-branes.}---We have two categories of D7-brane configurations at zero charge density: the Minkowski embeddings (MEs) and the black hole embeddings (BEs). MEs are the configurations of D7-branes without touching the black hole horizon, while BEs are those ending at the horizon. 
The quark-antiquark bound states are stable on MEs, while they are unstable on BEs~\cite{mesons}. Since the bound states are neutral, there is no charge carrier on MEs and the system behaves as an insulator. On the other hand, the bound states are unstable on BEs and will be broken into the charge carriers; the system has finite conductivity in this case. The location of the D7-brane with $J=E=0$ is governed by the current quark mass $m_{q}$ and the size of the horizon determined by $T$. The choice between the two embeddings is given by the competition between $m_{q}$ and $T$; MEs are thermodynamically favored at sufficiently large $m_{q}/T$, while BEs will be chosen at sufficiently small $m_{q}/T$. Therefore, the system undergoes a phase transition from insulator to conductor at the critical temperature determined by $m_{q}$. The system is a conductor in the high-temperature phase.

The situation changes if we add a finite external electric field $E$. The transition temperature will be affected by $E$. Furthermore, a sufficiently large $E$ creates a sufficiently large potential for the quarks and the antiquarks in comparison with their binding energy, and it will induce a breakdown of the insulation. In other words, there is an insulator-to-conductor transition at the critical electric field $E_{c}$ even in the low-temperature regime~\cite{Erdmenger:2007bn,Albash:2007bq}.

{\bf Numerical method.}---We have already chosen $\lambda=(2\pi)^{2}$ and $N=N_{c}N_{f}/(2\pi)^{2}$. Let us choose the temperature to be $T=\sqrt{2}/\pi$ so that $z_{H}=1$, $e=E/2$ and $z_{*}=\sqrt{E^{2}/4+1}-E/2$.\footnote{
In this Letter, we have employed the natural units $c=\hbar=k_{\mbox{\scriptsize B}}=1$. If our scale unit is meV (mili eV), $T\sim 5\ {\rm K}$. If we identify the unit quark charge with the unit charge of electrons, the effective fine-structure constant read from the Coulomb interaction in the inter-quark potential is $\sim 1$.} We further fix $N_{c}N_{f}=40$. 
$N_{c}N_{f}$ determines the pair-creation rate of the charge carriers as we shall explain later.
We need to solve the EOM for $\theta$ numerically. The natural boundary conditions are $\theta(z)/z|_{z=0}=m_{q}$ and $\theta^{\prime}|_{z=z_{H}}=0$, where the latter comes from the EOM at the horizon. However, we can also give the second condition at the singular shell~\cite{Albash:2007bq} instead. The condition for us is $\theta^{\prime}|_{z=z_{*}}=\cot\theta(z_{*})[B-\sqrt{B^{2}+C^{2}\tan^{2}\theta(z_{*})}]/(Cz_{*})$, where $B=3+2z_{*}^{4}+3z_{*}^{8}$ and $C=3(1-z_{*}^{8})$. This comes from the EOM at the singular shell with the assumption $\theta(z_{*})\neq\pi/2$.

We proceed as follows. First, we start with a given $E$ and compute $z_{*}$. Then, we assign $\theta(z_{*})$ as we like and compute $J$. At this stage, all the numerical factors in the EOM are assigned. Next, we solve the EOM for $\theta$ between the singular shell and the boundary of the AdS-BH by using the boundary conditions $\theta(z_{*})$ and $\theta^{\prime}(z_{*})$ specified above.
Then, we can read $m_{q}$ from $\theta(z)/z|_{z=0}$; we obtain $m_{q}$ as a function of $J$ and $E$.
Since the numerical analysis becomes unstable at the singular shell and the boundary, we introduce small positive cutoffs $\epsilon_{1}$, $\epsilon_{2}$, and the evaluations at the singular shell (boundary) are actually given at slightly outside (inside) of it: $z=z_{*}-\epsilon_{1}$ ($z=\epsilon_{2}$). Although we do not use the solutions inside the singular shell, we need to check the presence of the smooth solution until the horizon.

\begin{figure}[h]
    \centerline{\includegraphics[height=60mm]{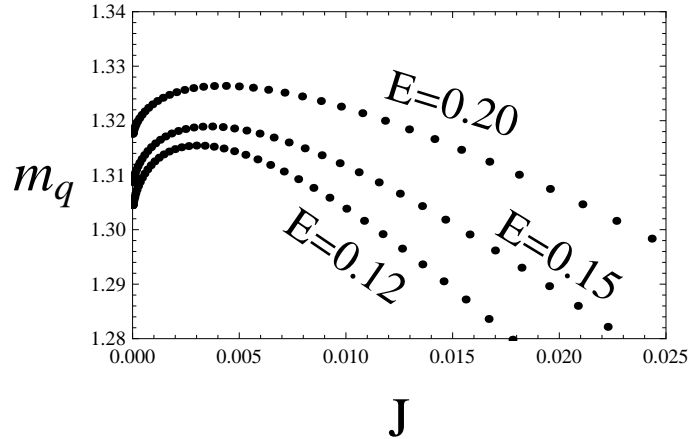}}
  \caption{$J$-$m_{q}$ curves at $E=0.12$, $0.15$, and $0.20$. $m_{q}$ is maximum at a nonzero but small value of $J$.}
  \label{fig:result1.eps}
\end{figure}

{\bf Results.}---Examples of $J$-$m_{q}$ curves at several values of $E$ are shown in Fig.~\ref{fig:result1.eps}. Of course, $m_{q}$ has a unique value at a given model and we need to choose some particular value of $m_{q}$. The typical $J$-$E$ characteristics of SCES are the S-shaped NDR, and the best strategy for us to find the NDR is to see whether we have different solutions at a common $E$ in the insulation-broken regime $E\ge E_{c}$ on BEs. 
We find that, indeed, there are two different solutions (for BEs) at given $m_{q}$ in some parameter region. Since they have different $J$'s at a common $E$, the system shows the double-valued nature of the conductivity. If we also count ME as a zero-conductivity branch, we may call it a triple-valued nature.  Furthermore, if we increase $E$ along the given $m_{q}$, the smaller $J$ {\em decreases} while the larger $J$ increases; the smaller-$J$ branch shows NDR, whereas the larger-$J$ branch has a positive differential resistivity.
\begin{figure}[h]
    \centerline{\includegraphics[height=60mm]{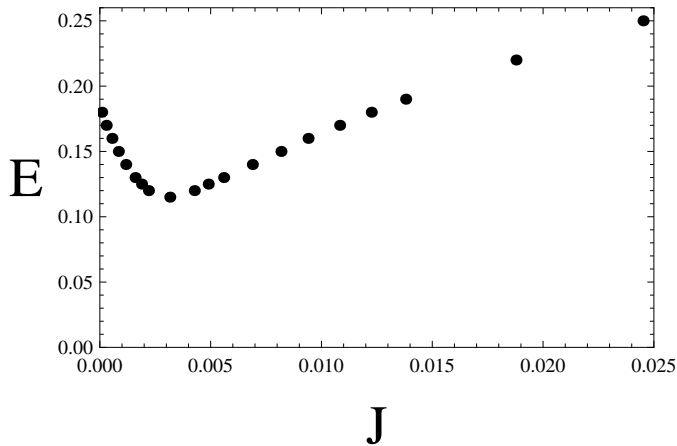}}
  \caption{$J$-$E$ curve at $m_{q}=1.315$. $E_{c}=0.11$ in this case. NDR appears in $J\le 0.0031$ and is absent for $E\ge 0.19$.}
  \label{fig:result2.eps}
\end{figure}
Note that the smaller-$J$ branch is a very narrow window in the full part of the $J$-$m_{q}$ curve. For example, the $J$-$m_{q}$ curve at $E=0.2$ extends until $J=0.288$, 
and the width of the smaller-$J$ branch along the $J$ axes is less than 2\% of the full part.
The detailed analysis shows that the highest value of $m_{q}$ approaches around $1.310$ at the $E\to +0$ limit, suggesting that $E_{c}=0$ if $m_{q}<1.310$. This is consistent with the fact that the system is a conductor at sufficiently small $m_{q}$ in comparison with $T$ (or sufficiently high $T$ in comparison with $m_{q}$). 

An example of $J$-$E$ relation at $m_{q}=1.315$ is given in Fig.~\ref{fig:result2.eps}.\footnote{If we choose our scale unit to be meV, the critical electric field $E_{c}$ in Fig.~\ref{fig:result2.eps} is $E_{c}\sim 5\times 10^{-1} {\rm V/m}$, and the current density realized at $E=E_{c}$ is $J\sim 1\times 10^{-4} {\rm mA/mm^{2}}$. } The system is an insulator for $E<E_{c}=0.11$. If $E\ge  E_{c}$, the insulation is broken and we observe a current. NDR is realized in the smaller-$J$ region. We always have the $J=0$ branch (ME branch) on top of the vertical axis. Therefore, our interpretation is that  Fig.~\ref{fig:result2.eps} shows the B-C-D region of Fig.~\ref{fig:S-shape} with the axes swapped; our NDR falls within the S-shaped NDR. There may be a small tunneling current that almost overlaps with the vertical axis, but we could not detect it within our numerical precision.\footnote{In the experimental data, shown in the inset of Fig.~1b in Ref.~\citen{exp1-2} for example, the current axis is given in the logarithmic scale and the tunneling current (which corresponds to the branch between A and B in Fig.~\ref{fig:S-shape}) almost overlaps with the zero-current axis in the linear scale.} We leave the detailed analysis on the tunneling current in a future work.

It is important to clarify what is the physically essential process in our NDR. Let us consider the doped cases. 
We can also ``dope'' the system by introducing finite quark-charge density~\cite{Nakamura:2006xk,Kobayashi:2006sb}. In this case, the D7-brane cannot take ME~\cite{Kobayashi:2006sb} and the system is always a conductor. The current is given by~\cite{Karch:2007pd}
\begin{eqnarray}
J=\sqrt{\sigma_{0}^{2}+d^{2}/(e^{2}+1)}\:E,
\label{Jatd}
\end{eqnarray}
where $d$ is related to the quark-charge density $\rho$ through $d=\rho/(\frac{\pi}{2}\sqrt{\lambda}T^{2})$. Owing to the doped charges, any small $E$ causes a current and we observe Ohm's law in the small-$J$ region. If we raise $J$, we may again observe NDR owing to the nontrivial behavior of $\sigma_{0}$.
It is indeed the case if $d$ is small enough not to smear the contribution of $\sigma_{0}$.
In this case, the curve in Fig.~\ref{fig:result2.eps} will be ``N-shaped'', (S-shaped in the sense of Fig.~\ref{fig:S-shape}) starting at the origin. The point is that the $d$-dependent term in the square root in (\ref{Jatd}) does not have any structure to produce NDR. Therefore, the $\sigma_{0}$-part in (\ref{Jatd}) is crucial for NDR. 
It is understood that the current due to the $\sigma_{0}$-part is caused by the pair creation of the charge carriers. The reasons are as follows: it contributes the current with the total system being kept neutral, and it vanishes if the mass of the charge carriers $m_{q}$ is infinite.~\cite{Karch:2007pd}\footnote{We also point out that $\sigma_{0}$ is proportional to $N_{c}N_{f}$. This suggests that it may be a one-loop contribution of the quarks, as in the perturbative computation of the pair-creation rate. Note that the quark loops have been taken into account to the 1-loop order in the probe approximation.}  As a conclusion, the pair-creation process is essential for our NDR.   

{\bf Discussion.}---Let us discuss the possible connections of our results to the realistic materials.  
Our system has the following features similar to those of the excitonic insulators.\footnote{A similarity between excitons and mesons has also been utilized in Ref.~\citen{Araki:2010gj}.} 1) The positive charges (``holes'') and the negative charges (``electrons'') are strongly correlated via the Coulomb-like interaction. 2) They form the neutral bound states (``excitons'') at low temperatures, and the system is an insulator (``excitonic insulator'') there. 3) The system exhibits insulator-conductor transition if we raise the temperature or the external electric field. 4) The interaction between the charges is mediated by the gluons (``phonons'' or ``photons'')\footnote{
It may be better to refer to neutral bipolarons rather than excitons if we compare the gluons with phonons. However, we are not strict in distinguishing them in this article. } that are neutral for the electric field.
If these properties are essential for NDR, we may also observe NDR in the realistic excitonic insulators right above the critical electric field of the insulation breaking. As far as the author knows, the nonlinear charge transport in the excitonic insulators have not yet been well explored either experimentally or theoretically. We suggest that experimental physicists see whether the NDR depicted in the present work is observed in some (candidates of) excitonic insulators.\footnote{
The system studied in Ref.~\citen{exp2} exhibits the neutral-ionic transition where the charge-transfer excitons may play a role. }

We can also see our results from the viewpoint of the quark-hadron physics. Let us consider the sQGP state~\cite{Gyulassy:2004zy} where the quark-antiquark bound state exists in the deconfinement phase of gluons. Our results suggest that the quarks are liberated at the critical value of the electric field and their current may show NDR. It is important to study how general this NDR is, in quark/meson systems and in the systems of charge-anticharge bound states.

We can suggest a phenomenological model of NDR. The phenomenological origins of NDR are classified into three types (except for the tunnel effect for some semiconductor junctions) in Ref.~\citen{book}: 1) nonlinearity of mobility, 2) nonlinearity of carrier density, and 3) nonlinearity of the electron temperature.\footnote{See also Ref.~\citen{Joule}.} We have found that our NDR originates in the pair creation of the charge carriers but not in the normal current of the doped charges. This means that the above feature 2) is crucial in our NDR. 
Although further study is necessary to reach the final conclusion, it is natural to assume that both the normal current and the pair-created current contribute 1) and 3) regardless of the origin of the charge carriers. If this assumption is right, 1) and 3) do not seem to be important in our NDR. 
The behavior of our NDR is in the category of the ``SNDC'' in Ref.~\citen{book} and it may be attributed to the impact ionization explained in Ref.~\citen{book}. The recent proposal of the many-body avalanche model of NDR\cite{exp2,O-K-A} also matches our picture. It is important to study further the connection between our results and the phenomenological models of NDR.
We expect that the present system is a good theoretical playground for studies on nonlinear charge transport and nonequilibrium steady states. 
The AdS/CFT correspondence can be a new tool for studying nonequilibrium physics as we have demonstrated here.

{\bf Acknowledgments.}---The author thanks S.~Ajisaka, H.~Aoki, H.~Hayakwa, Y.~Hidaka, H.~Kawai, T.~Kunihiro, H.~Ooguri, A.~Shimizu, H.~Suganuma, H.~Wada and especially T.~Oka for useful discussions. 
The author thanks the
Institute for the Physics and Mathematics of the Universe (IPMU), 
where this work was initiated during the Focus Week Activity ``Condensed Matter Physics Meets High Energy Physics.'' 
This work was supported by MEXT KAKENHI (21105006),
Grant-in-Aid for Scientific Research on Innovative Areas ``Elucidation of New Hadrons with a Variety of Flavors,''
and
the Grant-in-Aid for the Global COE Program ``The Next Generation of Physics, Spun from Universality and Emergence'' of MEXT of Japan.


%

\end{document}